\documentclass[aps,pra,preprint]{revtex4-1}
\usepackage[colorlinks,linkcolor=blue,urlcolor=black,citecolor=blue]{hyperref}
\usepackage{amsmath,amssymb,graphicx,subfigure,braket,siunitx,url,booktabs,multirow,appendix}
\usepackage{lipsum}
\begin{document}


\title{Quantum interference control of perfect photon absorption in a three-level atom-cavity system}

\author{Miaodi Guo}
\email{guomiaodi@xatu.edu.cn}
\affiliation{School of Sciences, Xi'an Technological University, Xi'an 710021, China}

\date{\today}

\begin{abstract}
We analyze a scheme for controlling coherent photon absorption by cavity electromagnetically induced transparency (EIT) in a three-level atom-cavity system. Coherent perfect absorption (CPA) can occur when time-reversed symmetry of lasing process is obtained and destructive interference happens at the cavity interfaces. Generally, the frequency range of CPA is dependent on the decay rates of cavity mirrors. When the control laser is settled, the smaller cavity decay rate causes the wider frequency range of CPA, and the input intensity is larger to satisfy CPA condition for a given frequency. While the cavity parameters are determined, Rabi frequency of the control laser has little effect on the frequency range of CPA. However, with EIT-type quantum interference, the CPA mode is tunable by the control laser. This means the CPA with given frequency and intensity of an input laser can be manipulated as the coherent non-perfect absorption (CNPA). Moreover, with the relative phase of input probe lasers, the probe fields can be perfectly transmitted and/or reflected. Therefore,  the system can be used as a controllable coherent perfect absorber or transmitter and/or reflector, and our work may have practical applications in optical logic devices. 
\end{abstract}

\maketitle

\section { \uppercase{introduction}}
The ability to control optical absorption especially to realize extreme absorption is crucial for the development of photonic devices \cite{Wang2017a,Trainiti2019,Horng2020} and optical quantum information processing, e.g., quantum communication and quantum computation \cite{Trautmann2015,Kang2020}. Electromagnetically induced transparency (EIT) \cite{Field1991}, which happens when the destructive interference of two transition path occurs, has been used for manipulating photon absorption in optical storage, quantum phase gate, optical switching, and optical probe \cite{Borges2016,Hao2019,Gao2019,Ann2020}. Generally, the EIT system is with a strong coupling laser and the strong nonlinearity. Even though, the absorption efficiency is less than $100\%$. Recently, studies show that the coherent perfect absorption (CPA) has captured significant attention in light-matter interaction, and can be realized in solid-state Fabry-Perot devices \cite{Chong2010,Page2011},  metamaterials \cite{Hu2016,Kang2018} and cavity quantum electrodynamics (CQED)  \cite{Wei2018a,Xiong2020}. Besides the manipulation on photon absorption, CPA has potential applications in photon detection and optical sensor \cite{Jeffers2019,Suwunnarat2019}. The physical origin behind CPA is the interference between transmitted and reflected fields at cavity interfaces. And the interaction of input fields and media must satisfy time-reversed-symmetry condition of radiation to realize CPA \cite{Zhang2012}.

Previous studies of CPA are in the linear classical regime or two-level-atom nonlinear quantum regime \cite{Agarwal2016,Rothenberg2016}, where the CPA mode is generally non-tunable. For example, in a two-level atom-cavity system \cite{Agarwal2016}, dual-frequency CPA modes at $\Delta_p=\pm4.5\Gamma$ are obtained for a given input intensity $|a_{in}|^2\approx55$ ($\Delta_p$ is the frequency detuning of probe field, and $\Gamma$ is the decay rate of atomic exucited level). The frequency of CPA is variable only with some certain intensities of probe laser. For practical applications, the frequency of CPA should be tunable for a given signal. Therefore, the media susceptibility of proposed scheme must be controllable to manipulate the interference at the interfaces. One of the methods is introducing EIT-type quantum interference in which a strong coherent laser couples the exucited state of a two-level atom with another fine level of ground state of the atom. 

Here, we propose a scheme where CPA modes are manipulated in a three-level-atom nonlinear quantum regime. In the scheme, some three-level atoms are trapped in a single-mode cavity which is driven by two coherent probe lasers from two ends. A control laser is coupled to the atoms causing two transition path with the destructive interference. And the interaction of input fields and atoms can be manipulated by the control laser. With the EIT-type quantum interference by control laser, CPA mode does not occur at resonant frequency of the probe field, instead, four CPA modes occur at non-resonant frequencies for a given input intensity of probe field. In addition, at a given frequency of probe field, CPA is transferred into coherent non-perfect absorption (CNPA) by control laser, and vise versa. The scheme provides the potential application of controllable perfect absorber in complicated system. 

In Sec.  \uppercase\expandafter{\romannumeral2}, we present the theoretical model, and we obtain the solution of output fields by master equation and input-output relation.  We also analyze the CPA condition and the manipulation of it by control laser. In Sec. \uppercase\expandafter{\romannumeral3}, we firstly discuss the intensity and frequency of input probe fields for CPA. And then, we discuss the output-input relation with different relative phase of input probe fields. Finally, we discuss the manipulation on CPA and CNPA by control laser.  And the conclusion is presented in Sec. \uppercase\expandafter{\romannumeral4}. 

\section{ \uppercase{theoretical model and analysis}}
\begin{figure}[!hbt]
	\centering
	\subfigure{\includegraphics[width=6.5cm]{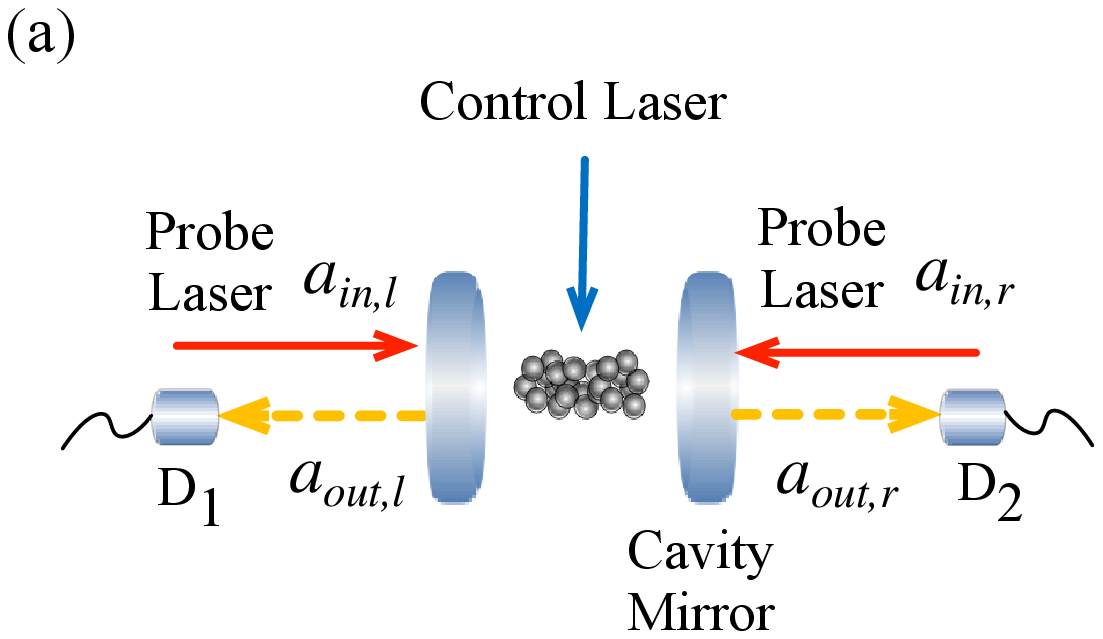}}
	\subfigure{\includegraphics[width=6.5cm]{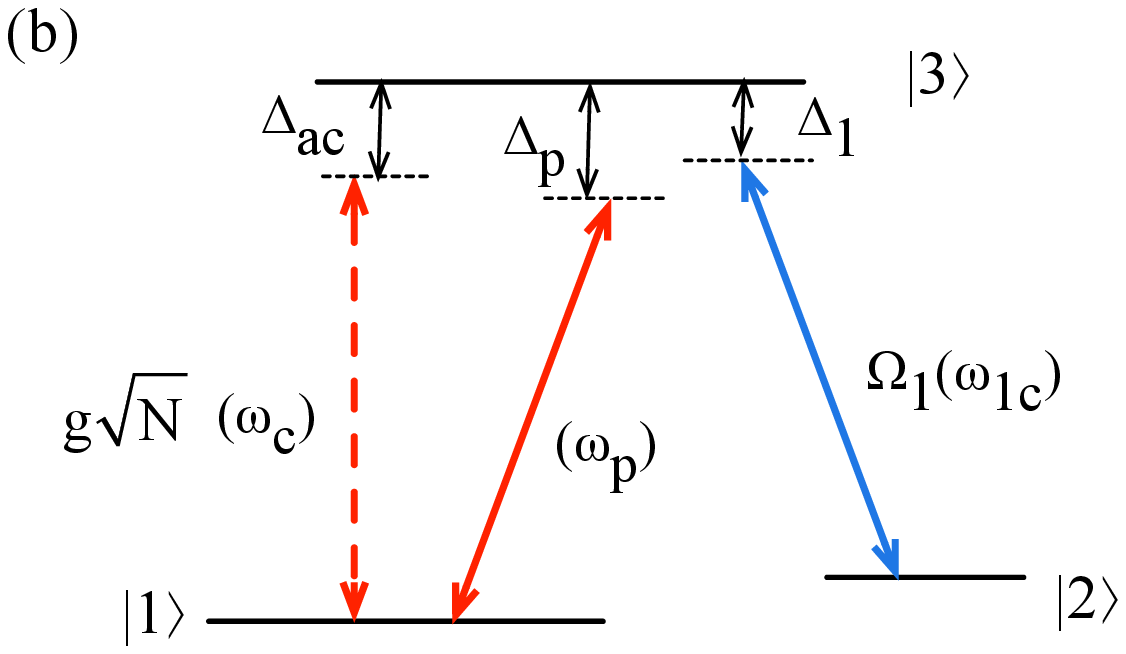}}
	\caption{(a) Schematic diagram of a two-sided cavity filling with some (b) three-level atoms.}\label{fig-system}
\end{figure}

The schematic diagram of the system is shown in Fig. \ref{fig-system}. We consider a system consisting of some three-level atoms coupled to a two-sided optical cavity. Two coherent probe lasers $a_{in,l}$ and $a_{in,r}$ are injected into the cavity from two interfaces, and drive the atomic transition $|1\rangle\rightarrow|3\rangle$ with a frequency detuning $\Delta_p=\omega_p-\omega_{31}$. $\Delta_{ac}=\omega_c-\omega_{31}$ is the frequency detuning between the cavity mode and the atomic transition $|1\rangle\rightarrow|3\rangle$. A control laser is coupled to the atomic transition $|2\rangle\rightarrow|3\rangle$ with a frequency detuning $\Delta_1=\omega_{1c}-\omega_{32}$. Here $\omega_c$ is the frequency of cavity mode, $\omega_p$ and $\omega_{1c}$ are frequencies of the probe and control lasers, and $\omega_{mn}$ ($m,n=1,2,3$) is the frequency of corresponding atomic transition. 

The Hamiltonian of the system is,
\begin{equation}
	\begin{split}
		H=&-\hbar  \sum _{j=1}^N [(\Delta_p-\Delta_1) \sigma_{22}^j+\Delta_p \sigma_{33}^j]-\hbar  (\Delta_p-\Delta_{ac})a^{\dagger}a\\
		&-\hbar  \sum _{j=1}^N (g a \sigma_{31}^j e^{-i\omega_p t}+\Omega_1 \sigma_{32}^je^{-i \omega_{1c}t})+H.C.\label{equ-H}
	\end{split}
\end{equation}
where $\sigma_{mn}^{j}=|m\rangle\langle n|$ ($m,n=1,2,3$) is the atomic operator, $a^{\dag}$($a$) is the creation (annihilation) operator of cavity photons, $g=\mu_{13}\sqrt{\omega_{c}/(2\hbar\varepsilon_{0}V)}$ is the cavity-QED coupling coefficient, and $\Omega_1=\mu_{23} E/\hbar$ is the Rabi frequency of the control laser. 

In semiclassical approximation, we treat the expectation values of field operators as the corresponding fields, e.g., $\langle a\rangle=\alpha$ ($\langle a^{\dagger}\rangle=\alpha^*$), and we treat density operator $\rho_{mn}$ as $\rho_{mm}=\langle \sigma_{mm}\rangle$, $\rho_{12}=\sigma_{12}e^{i(\omega_p-\omega_{1c})t}$, $\rho_{13}=\sigma_{13}e^{i\omega_p t}$, and $\rho_{23}=\sigma_{23}e^{i\omega_{1c}t}$ \cite{Sawant2016a}. The master equation with decay process can be written as, 
\begin{equation}
		\begin{split}
		\dot{\rho_{11}}&=\frac{\Gamma }{2} \rho_{33}+i g (\alpha^* \rho_{13}-\alpha  \rho_{31}),\\
		\dot{\rho_{12}}&= [i (\Delta_p-\Delta_1)-\gamma_{12}]\rho_{12}-i g \alpha  \rho_{32}+i \Omega_1 \rho_{13} ,\\
		\dot{\rho_{13}}&=(i \Delta_p-\frac{\Gamma}{2})\rho_{13} +i g \alpha (\rho_{11}-\rho_{33})+i \Omega_1\rho_{12} ,\\
		\dot{\rho_{22}}&=\frac{\Gamma}{2} \rho_{33}+i \Omega_1 (\rho_{23}-\rho_{32}),\\
		\dot{\rho_{23}}&= (i \Delta_1-\frac{\Gamma }{2})\rho_{23}+i g \alpha  \rho_{21}+i \Omega_1 (\rho_{22}-\rho_{33}),\\
		\dot{\rho_{33}}&=-\Gamma \rho_{33}+i g (\alpha  \rho_{31}-\alpha^* \rho_{13})+i \Omega_1 (\rho_{32}-\rho_{23}),\label{equ-density}
	\end{split}
	\end{equation}	
where $\Gamma$ is the decay rate of atomic level $|3\rangle$, and $\gamma_{12}$ is the decoherence rate between atomic levels $|1\rangle$ and $|2\rangle$. 

 The intracavity field can be solved from the following differential equation,
\begin{widetext}
\begin{equation}
	\dot{a}=i \left(\Delta _p-\Delta _{\text{ac}}\right)a+i g N \rho _{13}-\frac{\left(\kappa _l+\kappa _r\right)}{2}a+\sqrt{\frac{\kappa _l}{\tau }} a_{\text{in},l}+\sqrt{\frac{\kappa _r}{\tau }} a_{\text{in},r},\label{equ-dota}
	\end{equation}
\end{widetext}
where $\kappa_l$ ($\kappa_r$) is the field decay rate from left (right) cavity mirror, and $\tau$ is the photon round trip time inside the cavity. We consider a symmetric cavity with $\kappa_l=\kappa_r=\kappa$, the stable intracavity field $\alpha$ can be written as,
\begin{widetext}
\begin{equation}
	\alpha=\frac{\sqrt{\kappa/\tau} \left(\alpha _{\text{in},l}+\alpha _{\text{in},r}\right)}{\kappa-i \left(\Delta _p-\Delta _{\text{ac}}\right)-\frac{2  \Omega _1^2 g \alpha \left[2g^2 \left| \alpha \right|^2 \left(A-2 \gamma _{12} \Delta _p\right)+A\left(\Gamma\gamma _{12} +i \Gamma  \Delta _p+2 i \gamma _{12} \Delta _p-2 \Delta _p^2+2 \Omega _1^2\right)\right]}{\Gamma  \Omega _1^2 \left\{B+4\left[\gamma _{12}^2 \Delta _p^2+\left(\Delta _p^2-\Omega _1^2\right)^2\right]\right\}+g^2 \left| \alpha \right|^2 \left[C+8 \gamma _{12} \Omega _1^2 \left(\Delta _p^2+6 \Omega _1^2\right)\right]}},\label{equ-a}
	\end{equation}
\end{widetext}
where $A=\Gamma(\Delta_p+i\gamma_{12})$, $B=\Gamma ^2 \left(\gamma _{12}^2+\Delta _p^2\right)+4  \Gamma \gamma _{12} \Omega _1^2$, and $C=\Gamma ^3 \left(\gamma _{12}^2+\Delta _p^2\right)+8 \Gamma ^2 \gamma _{12}  \Omega _1^2+4 \Gamma  \Omega _1^2 \left(6 \gamma _{12}^2+4 \Delta _p^2+3 \Omega _1^2\right)$. Although the higher order items of $\alpha$ are neglected, it is clear that Eq. (\ref{equ-a}) is a cubic equation of $\alpha$. Therefore, according to the following input-output relations \cite{Agarwal2016,WallsDFandMilburnGJ2007}, 
\begin{equation}
	\begin{split}
	\langle a_{out,l}\rangle&=\sqrt{\kappa_l \tau}\langle a\rangle-\langle a_{in,l}\rangle,\\
	\langle a_{out,r}\rangle&=\sqrt{\kappa_r \tau}\langle a\rangle-\langle a_{in,r}\rangle,\label{equ-in-out}
	\end{split}
	\end{equation}
the output fields from two cavity mirrors are nonlinearly dependent on input probe fields, and are given as,
\begin{widetext}
\begin{equation}
	\begin{split}
	\frac{\left|\alpha_{out,l}\right|^2}{\left|\alpha_{in}\right|^2}&=\left|\frac{\kappa(e^{i\varphi}+1)}{\kappa-i \left(\Delta _p-\Delta _{\text{ac}}\right)-\frac{2  \Omega _1^2 g \alpha \left[2g^2 \left| \alpha \right|^2 \left(A-2 \gamma _{12} \Delta _p\right)+A\left(\Gamma\gamma _{12} +i \Gamma  \Delta _p+2 i \gamma _{12} \Delta _p-2 \Delta _p^2+2 \Omega _1^2\right)\right]}{\Gamma  \Omega _1^2 \left\{B+4\left[\gamma _{12}^2 \Delta _p^2+\left(\Delta _p^2-\Omega _1^2\right)^2\right]\right\}+g^2 \left| \alpha \right|^2 \left(C+8 \gamma _{12} \Omega _1^2 \left(\Delta _p^2+6 \Omega _1^2\right)\right)}}-e^{i\varphi} \right|^2,\\
	\frac{\left|\alpha_{out,r}\right|^2}{\left|\alpha_{in}\right|^2}&=\left|\frac{\kappa(e^{i\varphi}+1)}{\kappa-i \left(\Delta _p-\Delta _{\text{ac}}\right)-\frac{2  \Omega _1^2 g \alpha \left[2g^2 \left|\alpha\right|^2 \left(A-2 \gamma _{12} \Delta _p\right)+A\left(\Gamma\gamma _{12} +i \Gamma  \Delta _p+2 i \gamma _{12} \Delta _p-2 \Delta _p^2+2 \Omega _1^2\right)\right]}{\Gamma  \Omega _1^2 \left\{B+4\left[\gamma _{12}^2 \Delta _p^2+\left(\Delta _p^2-\Omega _1^2\right)^2\right]\right\}+g^2 \left|\alpha\right|^2 \left(C+8 \gamma _{12} \Omega _1^2 \left(\Delta _p^2+6 \Omega _1^2\right)\right)}}-1\right|^2.\label{equ-aout}
	\end{split}
	\end{equation}
\end{widetext}
where $\left|\alpha\right|^2$ can be solved by Eq. (\ref{equ-a}), and input probe fields are assumed as $a_{in,l}=\left|a_{in}\right|e^{i\varphi}$ and $a_{in,r}=\left|a_{in}\right|$. 

According to Eq. (\ref{equ-in-out}), the input fields must be identical when this system is acted as a perfect absorber, i.e., $a_{in,l}=a_{in,r}$ and $\varphi=0$. Therefore, from $\langle a_{out,l}\rangle=\langle a_{out,r}\rangle=0$, the intracavity field intensity with CPA condition can be derived as, 
\begin{widetext}
\begin{equation}
	\left|\alpha\right|^2=\frac{2 g^2 N \left(B-2 \gamma _{12} \Gamma  \Omega _1^2\right)-\Gamma  \kappa  \left\{B+4\left[\gamma _{12}^2 \Delta _p^2+\left(\Delta _p^2-\Omega _1^2\right)^2\right]\right\}}{g^2/\Omega _1^2 \left\{\kappa \left[C+8 \gamma _{12} \Omega _1^2 \left(\Delta _p^2+6 \Omega _1^2\right)\right]-4 \gamma _{12} \Gamma  g^2 N \Omega _1^2\right\}}.\label{equ-cpaa}
	\end{equation}
\end{widetext}
Considering Eqs. (\ref{equ-a}) and (\ref{equ-cpaa}) together, the input intensity and frequency to satisfy CPA condition can be derived, however the solutions are not presented here because of their complicated analytic forms. Even though, according to Eqs. (\ref{equ-in-out}) and (\ref{equ-cpaa}), the relation between input intensity and frequency of CPA can be derived as the following simple form, $\left|\alpha_{in}\right|^2=T \left|\alpha\right|^2$, where $T=\kappa \tau$ is the transmission coefficient of the cavity. It is clear that the formula is a quartic equation about $\Delta_p$, therefore, varying the system parameters to match the CPA condition can lead to the manipulation on single-, dual-, three- and four-frequency CPA.

When $\Omega_1=0$, the scheme can be acted as a two-level CPA system \cite{Agarwal2016}. While $\Omega_1\neq 0$, the coupling between atomic level $|2\rangle$ and $|3\rangle$ forms a destructive quantum interference of two absorption channels of transition $|1\rangle\rightarrow|3\rangle$. And this will induce the interference control on CPA condition. In the following, we will study the manipulation on CPA from decay rates ($\kappa$) of cavity mirrors and the Rabi frequency ($\Omega_1$) of the control laser. Furthermore, we also study the output intensity controlled by the frequency, the intensity, and the phase of the input probe laser. 

\section{\uppercase{numerical results and discussion}}
\begin{figure}[!hbt]
	\centering
	\subfigure{\includegraphics[width=6cm]{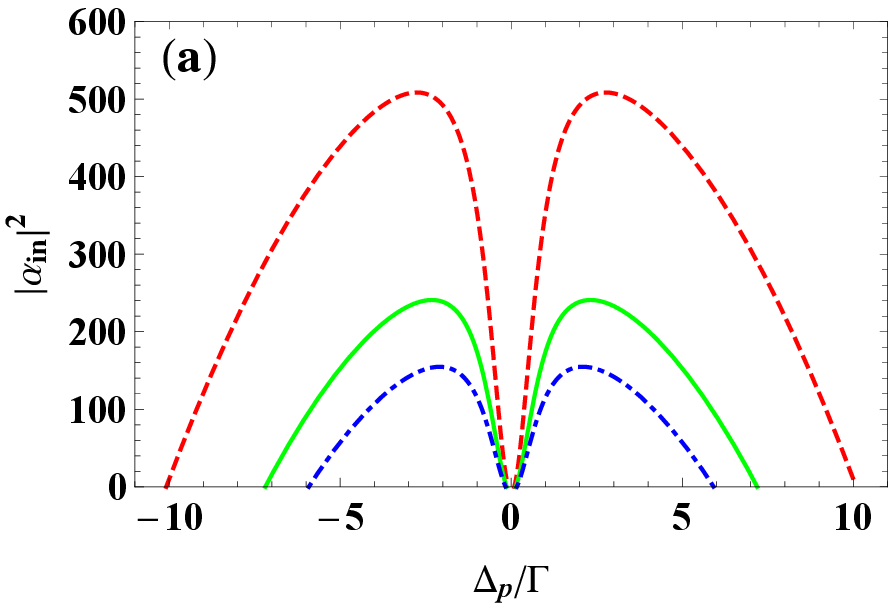}}
     \subfigure{\includegraphics[width=6cm]{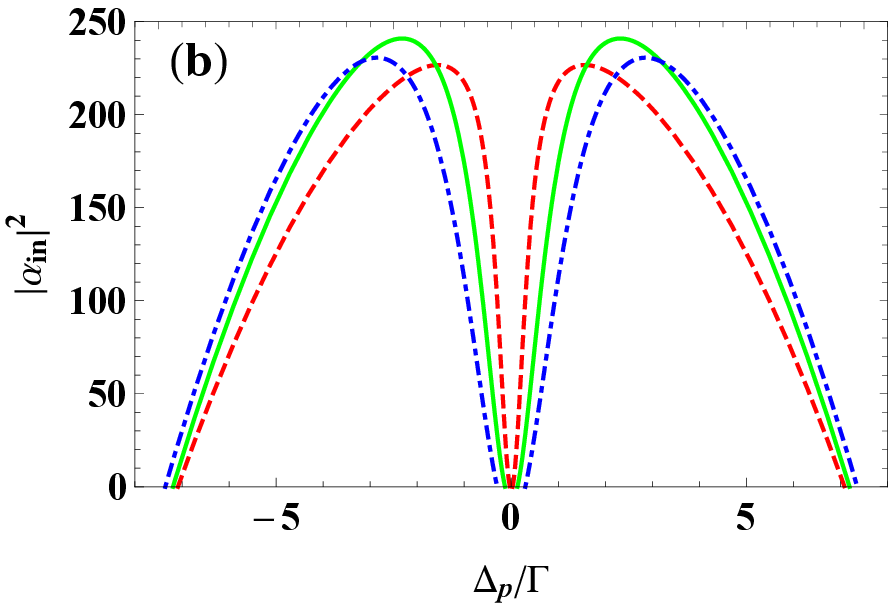}}
     \caption{Input intensity of CPA for (a) $\Omega_1=\Gamma$ with $\kappa=0.5\Gamma$ (dashed red line), $\kappa=\Gamma$ (solid green line), and $\kappa=1.5\Gamma$ (dot-dashed blue line), and (b) $\kappa=\Gamma$ with $\Omega_1=0.5\Gamma$ (dashed red line), $\Omega_1=\Gamma$ (solid green line), and $\Omega_1=1.5\Gamma$ (dot-dashed blue line).}\label{fig-cpaain}
	\end{figure}

Fig. \ref{fig-cpaain} shows the frequency range of CPA. The parameters in this section are, $\gamma_{12}=0.001\Gamma$, $g\sqrt{N}=10\Gamma$, and $T=0.01$. In Fig. \ref{fig-cpaain}(a), for EIT system, the output intensity can be zero at resonant frequency  only when input probe field is absent. This is caused by zero absorption of atomic media at resonant frequency, which causes $n^{''}=0$ ($n^{''}$ is the imaginary part of refractive index). Together with the quartic equation of $\left|\alpha_{in}\right|^2$ and $\Delta_p$, it accounts for the phenomenon shown in Fig. \ref{fig-cpaain}, where four-frequency CPA occurs at a specific input intensity in the three-level cavity-EIT system rather than the dual-frequency CPA in the two-level atom-cavity system \cite{Agarwal2016}. While at non-resonant frequencies, the input intensity is stronger to satisfy CPA condition for smaller decay rate $\kappa$, which also forms the wider frequency range of CPA. This indicates that a higher cavity finesse leads to a stronger nonlinear excitation regime of CQED.

In Fig. \ref{fig-cpaain}(b), the frequency range of CPA has little difference for different $\Omega_1$. However, for a specific frequency of the probe field, the intensity to realize CPA is usually different for different $\Omega_1$.  For example, when $\left|\Delta_p\right|=\pm5\Gamma$, the needed input intensities are $\left|\alpha_{in}\right|^2=125.5$ for $\Omega_1=0.5\Gamma$, $\left|\alpha_{in}\right|^2=152.9$ for $\Omega_1=\Gamma$, and $\left|\alpha_{in}\right|^2=165.8$ for $\Omega_1=1.5\Gamma$, respectively. This is on account of the manipulation on polariton states by the control laser. In the system, the first exucited polariton states contain two bright polariton states $|\Psi_+\rangle=\frac{1}{\sqrt{2}}(\frac{1}{\sqrt{N}}\sum_{j=1}^{N}|1,...3_{j},...1\rangle|0_c\rangle+|1,...1,...1\rangle|1_c\rangle)$ and $|\Psi_-\rangle=\frac{1}{\sqrt{2}}(\frac{1}{\sqrt{N}}\sum_{j=1}^{N}|1,...3_{j},...1\rangle|0_c\rangle-|1,...1,...1\rangle|1_c\rangle)$, and a dark polariton state $|\Psi_d\rangle=\frac{1}{\sqrt{g^2N+\Omega_1^2}}(g\sum_{j=1}^{N}|1,...2_{j},...1\rangle|1_c\rangle-\Omega_1|1,...1,...1\rangle|0_c\rangle)$. Here $|1_c\rangle$ and $|0_c\rangle$ are one-photon and zero-photon states of the cavity mode. With $\Omega_1 \ll g\sqrt{N}$, the control laser can be treated perturbatively, and then the bright polariton state $|\Psi_+\rangle$ will be split into two dressed polariton states $|\Phi_+\rangle=\frac{1}{\sqrt{2}}(|\Psi_+\rangle+|2\rangle)$ and  $|\Phi_-\rangle=\frac{1}{\sqrt{2}}(|\Psi_+\rangle-|2\rangle)$. The destructive interference between two excitation paths induces cavity EIT at the frequency of polariton resonance. Therefore, input probe fields are not absorbed by atoms. At other frequencies, the EIT-type quantum interference makes the output intensities from two interfaces dependent on the relative phase $\varphi$ and intensities of the probe fields.

\begin{figure}[!hbt]
	\centering
	\subfigure{\includegraphics[width=6cm]{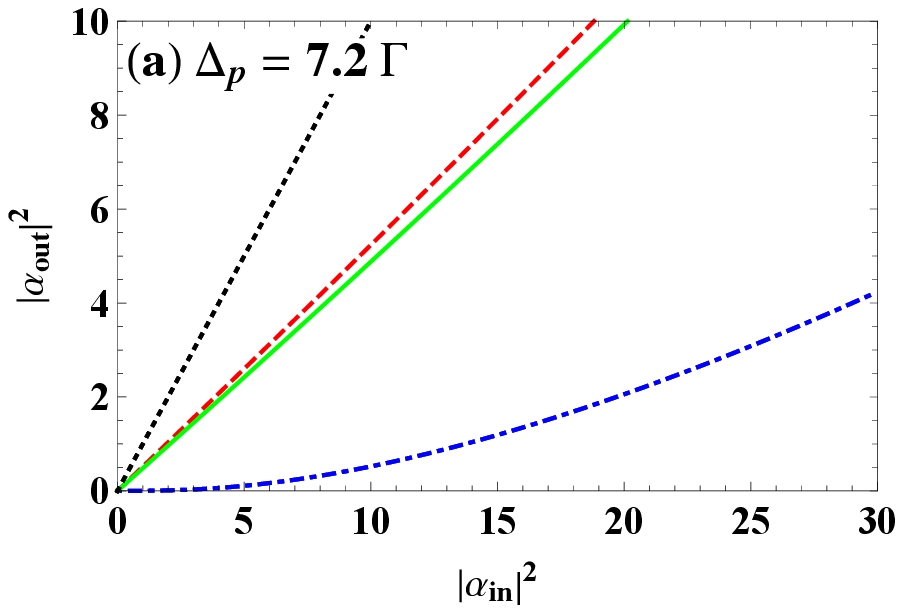}}
	\subfigure{\includegraphics[width=6cm]{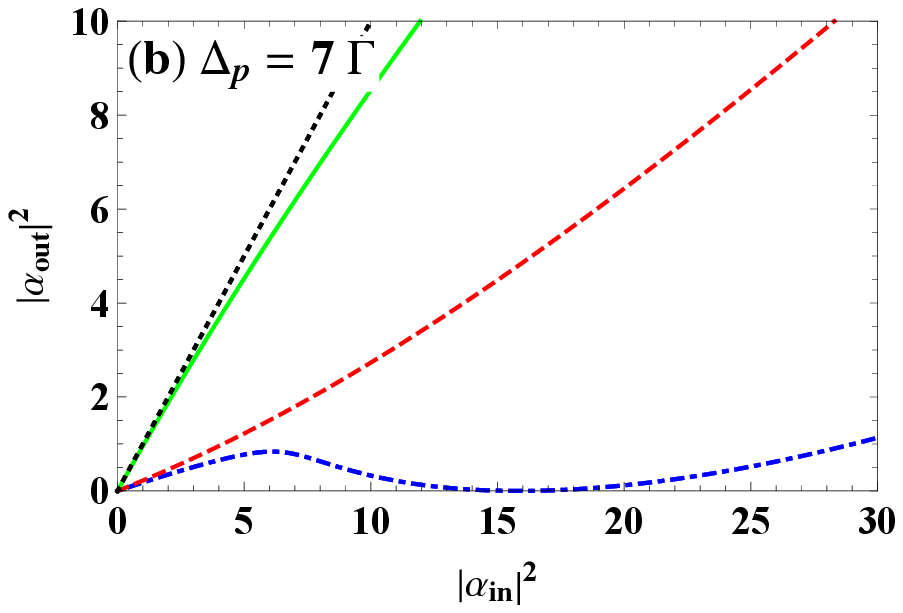}}
	\caption{The output intensity versus input intensity for (a) $\Delta_p=7.2 \Gamma$ and (b) $\Delta_p=7 \Gamma$ with $\varphi=0$ (dot-dashed blue lines), $\varphi=\pi/2$ (dashed red lines for $\left|\alpha_{out,r}\right|^2$ and solid green lines $\left|\alpha_{out,l}\right|^2$), and $\varphi=\pi$ (dotted black lines).}\label{fig-aout}
\end{figure}
In the following, we compare the difference of output intensity from two frequencies of input probe fields for $\varphi=0$, $\varphi=\pi/2$, and $\varphi=\pi$ in Fig. \ref{fig-aout}. When $\varphi=2n\pi$, CPA is manifested at a given frequency and intensity of the probe laser because of the fully destructive interference of the transmission and the reflection of two input probe fields at two ends of the cavity. When $\varphi=(2n+1)\pi$, two output intensities $\left|\alpha_{out,r}\right|^2$ and $\left|\alpha_{out,l}\right|^2$ are identical and equal to the input intensity $|\alpha_{in}|^2$, the reason of which is that no light can be coupled into the cavity for the destructively interference of two input probe fields \cite{Agarwal2016}. At other values of $\varphi$, two output intensities $\left|\alpha_{out,r}\right|^2$ and $\left|\alpha_{out,l}\right|^2$ are generally different, and CPA does not occur.

\begin{figure}[!hbt]
	\centering
	\subfigure{\includegraphics[width=6cm]{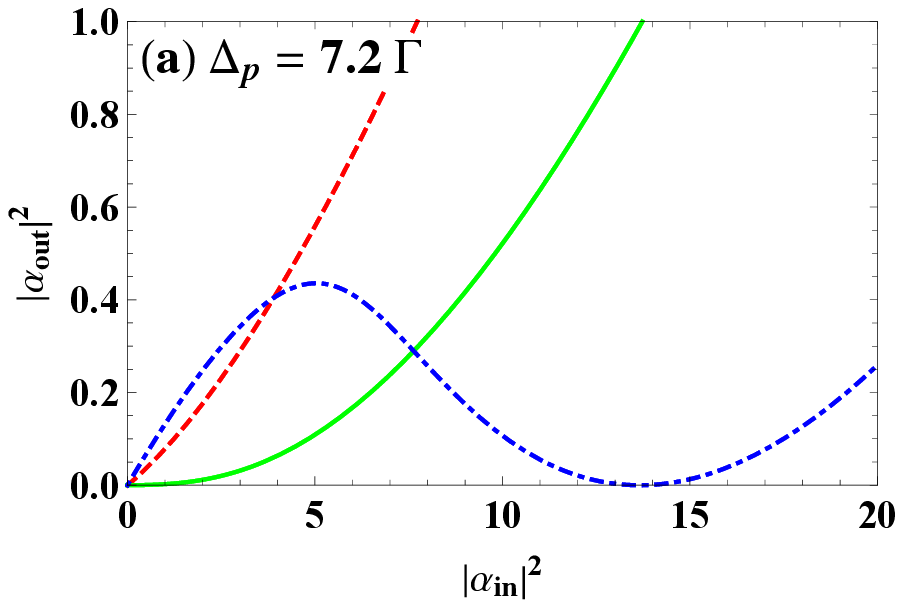}}
	\subfigure{\includegraphics[width=6cm]{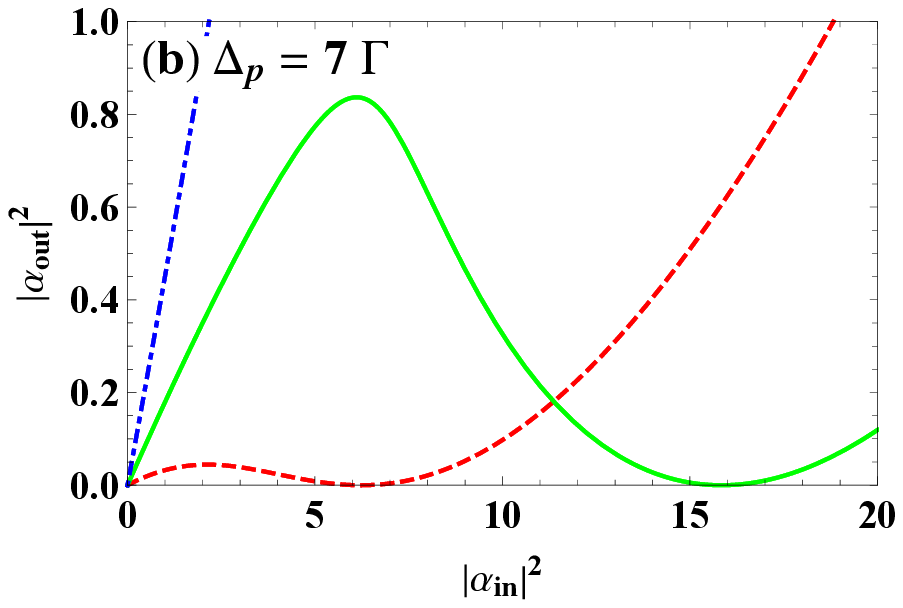}}
	\caption{The output intensity as the function of $\left|\alpha_{in}\right|^2$ with $\Omega_1=0.5\Gamma$ (dashed red lines),  $\Omega_1=\Gamma$ (solid green lines), and $\Omega_1=1.5\Gamma$ (dot-dashed blue lines).}\label{fig-cpa-aout}
\end{figure}

At a given frequency of input probe field, e.g., $\Delta_p=7.2\Gamma$, when control field $\Omega_1=\Gamma$, CPA does not occur for $\varphi=0$ as shown in Fig. \ref{fig-aout}(a). However, changing $\Omega_1$ to control single-path absorption can realize CPA, for example, when $\Omega_1=1.5\Gamma$, CPA can be manifested at the frequency (the dot-dashed blue line shown in Fig. \ref{fig-cpa-aout}(a)). In three-level EIT system, the manipulation on single-path absorption by $\Omega_1$ will also disturb CPA condition. For example, in Fig. \ref{fig-cpa-aout}(b), when $\Omega_1=0.5\Gamma$, the input intensity for CPA is 6.3. When $\Omega_1=\Gamma$, the needed intensity is 15.9. When $\Omega_1$ is increased as $\Omega_1=1.5\Gamma$, the needed intensity is increased as 30.3 which is not manifested with dot-dashed blue line in Fig. \ref{fig-cpa-aout}(b).

This indicates that for three- or four-level atom-cavity system, the coherent perfect absorption is realizable and controllable by EIT-type quantum interference. 

\section{\uppercase{conclusion}}
In conclusion, we have analyzed the interference control on CPA condition in a three-level atom-cavity system. Generally, the decay rates of the cavity mirrors can be as small as possible to widen the frequency range of CPA. While for an input probe laser with a given frequency, the needed input intensity of CPA is larger as the decay rates are smaller, which forms the CPA with a stronger excitation regime. However, the needed intensity of CPA is controllable by the control laser, which provides the application of CPA with weak fields in a low-loss cavity. Besides, a CPA mode is transferred to a CNPA mode by the control laser.  In addition, with EIT-type quantum interference induced by the control laser, four CPA modes instead of two CPA modes are obtained for a given input probe laser, which has potential application in the study of wide-band perfect absorber.

\section*{\uppercase{acknowledgment}}
We acknowledge support from Natural Science Foundation of Shaanxi Provincial Department of Education (Grant No. 20JK0682)

\bibliography{mybib}

\end{document}